\documentclass[10pt, aps, prb, superscriptaddress, noshowkeys, twocolumn,
floatfix, preprintnumbers,showpacs]{revtex4-1}

\usepackage{amsmath, amssymb, amsfonts}
\usepackage{subfigure}
\usepackage{graphicx, color}
\usepackage{times}

\usepackage{graphicx}
\usepackage{bm,color}
\usepackage{ulem}

\usepackage[breaklinks]{hyperref}
\usepackage{bookmark} 
\newcommand{\bk}{{\bf{k}}}
\newcommand{\bp}{{\bf{p}}}
\newcommand{\bq}{{\bf{q}}}

\newcommand{\MoTe}{{MoTe$_\mathrm{2}$-H}}

\newcommand{\bA}{{\bf{A}}}

\newcommand{\br}{{\bf{r}}}

\newcommand{\epspar}{\epsilon_{\parallel}}

\newcommand{\epsz}{\epsilon_{\bot}}
\newcommand{\rome}{{\mathrm{e}}}
\newcommand{\romh}{{\mathrm{h}}}
\newcommand{\romc}{{\mathrm{c}}}
\newcommand{\romv}{{\mathrm{v}}}

\newcommand{\romA}{{\mathrm{A}}}
\newcommand{\romB}{{\mathrm{B}}}

\allowdisplaybreaks
\draft 

\begin{document}
\title{{Ultrafast band-gap renormalization and build-up of optical gain in monolayer MoTe$_2$}}

\author{L. Meckbach}
\affiliation{Department of Physics and Material Sciences Center,
Philipps University Marburg, Renthof 5, D-35032 Marburg, Germany}

\author{J. Hader}
\affiliation{Wyant College of Optical Sciences, University of Arizona, 
1630 E. University Blvd., Tucson, AZ, 85721, USA}

\author{U. Huttner}
\affiliation{Department of Physics and Material Sciences Center,
Philipps University Marburg, Renthof 5, D-35032 Marburg, Germany}

\author{J. Neuhaus}
\affiliation{Department of Physics and Material Sciences Center,
Philipps University Marburg, Renthof 5, D-35032 Marburg, Germany}

\author{J.T. Steiner}
\affiliation{Department of Physics and Material Sciences Center,
Philipps University Marburg, Renthof 5, D-35032 Marburg, Germany}

\author{T. Stroucken\footnote{Author to whom correspondence should be addressed: tineke.stroucken@
physik.uni-marburg.de}}
\affiliation{Department of Physics and Material Sciences Center,
Philipps University Marburg, Renthof 5, D-35032 Marburg, Germany}

\author{J.V. Moloney}
\affiliation{Wyant College of Optical Sciences, University of Arizona, 
1630 E. University Blvd., Tucson, AZ, 85721, USA}

\author{S.W. Koch}
\affiliation{Department of Physics and Material Sciences Center,
Philipps University Marburg, Renthof 5, D-35032 Marburg, Germany}

\begin{abstract}
The dynamics of band-gap renormalization and gain build-up in monolayer \MoTe\ is investigated by evaluating the non-equilibrium Dirac-Bloch equations with the incoherent carrier-carrier and carrier-phonon scattering treated via quantum-Boltzmann type scattering equations. For the case where an approximately $300$\,fs-long high intensity optical pulse generates charge-carrier densities in the gain regime, the strong Coulomb coupling leads to a relaxation of excited carriers on a few fs time scale. The pump-pulse generation of excited carriers induces a large band-gap renormalization during the time scale of the pulse. Efficient phonon coupling leads to a subsequent carrier thermalization within a few ps, which defines the time scale for the optical gain build-up energetically close to the low-density exciton resonance.
\end{abstract}
\date{14 January 2020}


\maketitle 


\section{Introduction}
Monolayers (MLs) of transition-metal dichalcogenides (TMDCs) hold great promise
as active material in next generation opto-electronic devices. Unlike their bulk counterparts, MLs of many semiconducting TMDCs exhibit a direct gap with transition energies in the visible to near-infrared regime\,\cite{mak2010,splendiani2010,Zhang2014,ding2011,qiu2013,novoselov2016,cheiwchanchamnangij2012}. As compared to conventional semiconductors, they provide strong light-matter coupling and many-body effects due to carrier confinement and weak intrinsic screening of the Coulomb interaction. At low excitation levels, the electron-hole attraction leads to the formation of excitons with large binding energies that absorb as much as $10$--$20$\,\% of the incoming light for a single layer\,\cite{chernikov2014,he2014,ye2014,zhu2014}. Because of this strong light-matter interaction, TMDC based photonic devices promise high efficiency and have the potential for saturable absorbers, nanoemitters or  nanolasers with the smallest possible amount of optically active material. Indeed, room temperature lasing has been reported for different TMDC materials for comparatively low pump intensities and emission frequencies centering around the respective A exciton resonances\,\cite{salehzadeh2015,yeyu2015,li2017}.

One of the key properties for operation and design of nano-photonic devices is the quasi-particle or optical band gap. Due to Coulombic renormalizations, the quasi-particle gap is modified by the presence of excited carriers and depends on the precise excitation conditions. In a conventional semiconductor where screening is strong, these band-gap renormalizations are typically in the meV range. In contrast, in TMDCs excitation induced band-gap shrinkages of several hundred meV have been reported in experimental\,\cite{chernikov2015,ulstrup2016} and theoretical\,\cite{steinhoff2014,meckbach2018b} investigations. The injection of external charge carriers has been proposed as a possibility to dynamically control the optical gap on a femtosecond time-scale\,\cite{chernikov2015,ulstrup2016}. Furthermore, carrier-carrier and carrier-phonon scattering lead to excitation induced dephasing and the build-up of screening, thus dynamically modifying the exciton binding and peak gain positions. In particular, for laser applications precise predictions for the peak gain are desirable to design optical cavities correspondingly.

In this paper, we use the example of \MoTe\ to perform a microscopic calculation of the carrier dynamics and optical gain development after non-resonant optical excitation.  Among semiconducting TMDC materials, MoTe$_2$-H provides the most favourable conditions to achieve optical plasma gain. Whereas in W-based TMDCs the fundamental gap corresponds to spin-forbidden, dark transitions, in MoTe$_2$-H, for each spin component, the fundamental gap is undoubtly direct 
with a relatively large spin splitting and offset between the side and global minima  in the conduction band.
Without such an offset substantial amounts of electrons can leak quickly from the K/K'-points to the side valley. This reduces the carrier inversion at the global band minima and reduces or even prevents optical gain.

A well established scheme to deduce the carrier dynamics and its influence on the optical spectra is to probe the optical response of the system at different delay times after excitation with a strong optical pump pulse. To simulate such a scenario, we extend our recently developed Dirac-Bloch equation (DBE) scheme\,\cite{stroucken2017,meckbach2018a,meckbach2018b,meckbach2018c} beyond the linear low-excitation and quasi-equilibrium regime. In particular, we include incoherent interactions due to electron-electron and electron-phonon scattering in order to study the carrier dynamics and to determine the dephasing of the optical polarizations and the resulting broadening of optical spectra self-consistently.

\section{Methods}
To compute the carrier dynamics and its influence on the optical spectra,
we use a hybrid density functional theory (DFT) and equation of motion (EOM) approach. In a first step, we determine the relevant  material parameters, i.e. the band structure, dipole-matrix elements, as well as the dielectric constants of bulk \MoTe , via density functional theory.
In order to compute the carrier dynamics and evolution of the optical spectra on a dense $\bk$-grid, we use an effective four band Hamiltonian that is based on the single-particle band structure and dipole-matrix elements derived from our DFT calculations to describe the regions of the Brillioun zone that are actually populated. Subsequently, we derive the EOM for interband polarizations that couple directly to the optical field, and the respective occupation probabilities of the involved bands. Since DFT-based band structure calculations usually underestimate the quasi-particle gap, we compute the ground-state band-gap renormalization self-consistently from the gap equations.

\subsection{DFT calculations }\label{subsec:DFT}

\begin{table*}[t!]
\caption{Material parameters for \MoTe , i.e. the conduction-band (valence-band) valley minima (maxima) $\epsilon^{\romc}_{K}$ and $\epsilon^{\romc}_{\Sigma}$  ($\epsilon^{\romv}_{K}$), effective masses $m^{*}_{K}$ and $m^{*}_{\Sigma}$, dipole-matrix elements $d^\pm_{K}$, as well as the dielectric constants $\epspar^B,\epsz$ and out-of-plane lattice constant $D\,(c/2)$, based on our DFT calculations. For the $K'$ and $\Lambda$ valleys, the spin components are interchanged. }
\begin{ruledtabular}
\begin{tabular}{cccccccccc}
Spin&$\epsilon^{\romc}_{K}$\,[eV]&$\epsilon^{\romv}_{K}$\,[eV]&$\epsilon^{\romc}_{\Sigma}$\,[eV]&$m^{*}_{K}$\,[$m_0$]&$m^{*}_{\Sigma}$\,[$m_0$]&$d^\pm_{K}$\,[e\AA]&$\epspar^B$&$\epsz$&$D\,(c/2)$\,[\AA]\\ \hline
$\uparrow  $&1.017&  0.0 &1.114&0.607&0.407&3.51&    &     &     \\
$          $&     &      &     &     &     &    &20.30&10.90&6.99\\
$\downarrow$&1.052&-0.214&1.099&0.728&0.428&2.88&    &     &     \\
\end{tabular}
\end{ruledtabular}
\label{tab:DFTparameters}
\end{table*}

\begin{figure}[t!]
\includegraphics[width =0.41\textwidth]{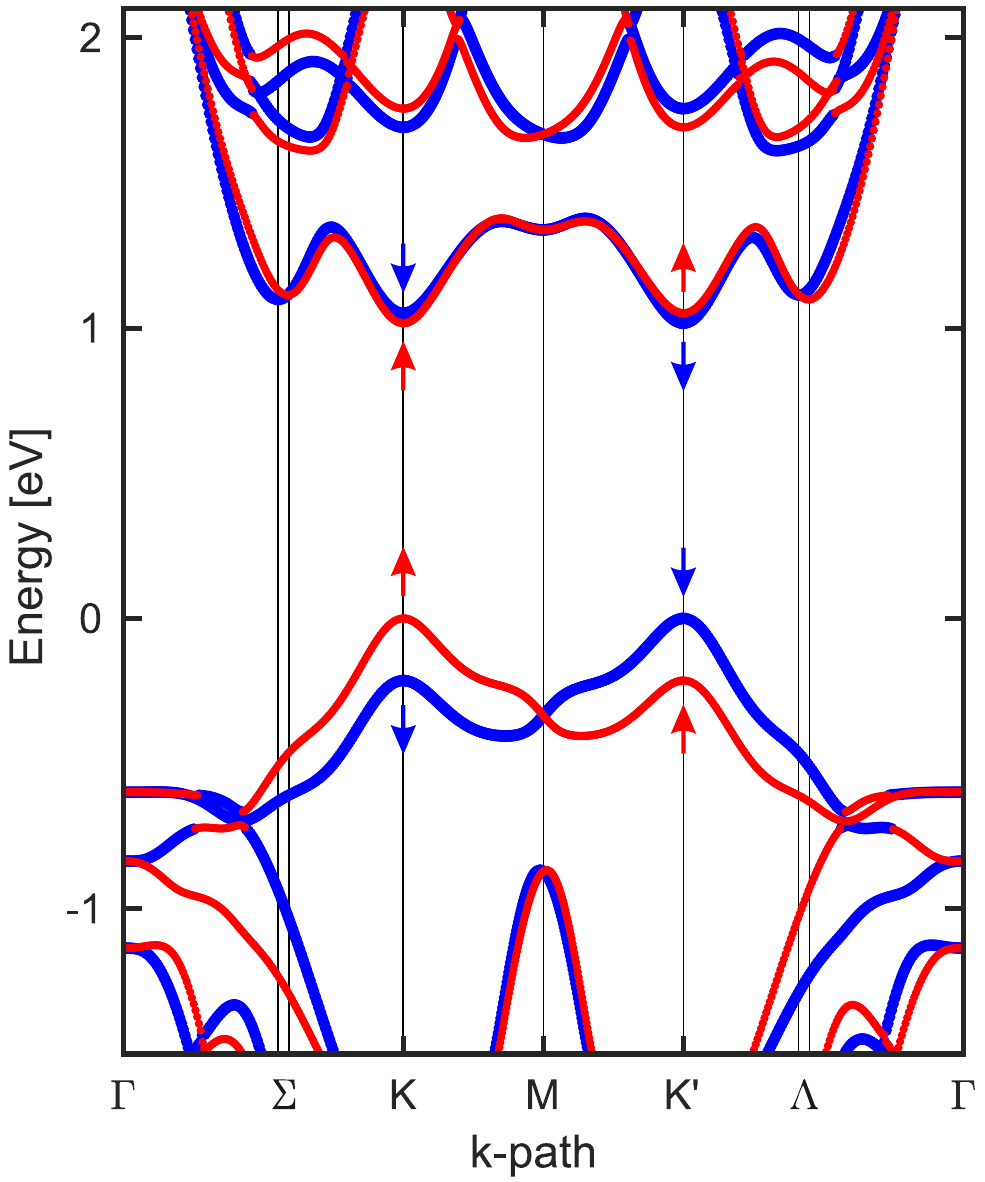}
\caption{\label{DFTbandstructure} DFT band structure of a \MoTe\ ML. Bands with spin up are depicted in red, bands with spin down in blue.}
\end{figure}

The relevant material parameters
are calculated via density functional theory (DFT)\,\cite{Kohn_PR140_1965} using the Vienna \emph{Ab initio} Simulation Package (VASP)\,\cite{Kresse_PRB47_1993,Kresse_PRB49_1994,Kresse_PRB54_1996,Kresse_CMS6_1996} and listed in Table\,\ref{tab:DFTparameters}. All computations employ the generalized-gradient-approximation via the Perdew-Burke-Ernzerhof (PBE) functional\,\cite{Perdew_PRL77_1996}, including the spin-orbit interaction\,\cite{Steiner2016}. The unit cell describing a \MoTe\ ML contains three atoms in total, while a vacuum region of $20$\,\AA{} around the ML is sufficient to prevent unphysical interactions with its periodic copies. The unit cell of bulk \MoTe\ in the common 2H form, used in the computations of the bulk dielectric constant, consists of two MLs and contains $6$ atoms. The Van-der-Waals interaction between neighboring layers is modeled via Grimme's dispersion correction method (PBE-D3)\,\cite{Grimme_JCP132_2010,Grimme_JCC32_2011}. In both cases, bulk and ML, a full relaxation of atomic positions and the unit cell shape and size is performed until all inter-atomic forces are smaller than $2.5\times10^{-3}$\,eV/\AA{}. The reciprocal space is sampled by a $15 \times 15 \times 3$ Monkhorst-Pack\,\cite{Monkhorst_PRB13_1976} $k$-mesh in the case of the ML and $10 \times 10 \times 10$ in the bulk case. The cutoff energy of the plane wave expansion is set to $750$\,eV for the structural relaxations and the MLs properties, while a value of $500$\,eV is used in the bulk case. The self-consistency cycle of the electronic minimization is repeated until an energy convergence criterion of $10^{-8}$\,eV is reached.

The resulting ML band structure is  shown in Fig.\,\ref{DFTbandstructure} and exhibits direct gaps at the $K$ and $K'$ points of the Brillioun zone with a non-interacting gap of $\Delta_\romA=1.017$\,eV and $\Delta_\romB=1.266$\,eV for the A ($K_\uparrow/K'_\downarrow$) and B ($K_\downarrow/K'_\uparrow$) bands. As in other TMDC materials, the atomic orbitals predominantly contibuting to the valence and conduction bands at the $K$ and $K'$ point are the $d$-type Mo-orbitals with equal parity. Furthermore, the conduction bands display  a spin splitting of $-35$ meV and side valleys at the $\Sigma/\Lambda$ points, that are $97$\,meV ($\Sigma_\uparrow/\Lambda_\downarrow$) and $82$\,meV  ($\Lambda_\uparrow/\Sigma_\downarrow$) above the respective $K/K'$-valley minima. These values are on the lower end of the range of published values that have been obtained using different functionals for the exchange correlation potential or  GW corrections\cite{ molinasanchez2015, kormanyos2015, novko2019} and sufficiently large to prevent an excitation induced transition from a direct to indirect band gap
\,\cite{erben2018,lohof2019}. 

The interband dipole-matrix elements are accessed via the linear optics routine in VASP as described in Ref.\,\onlinecite{Gajdos2006} and include contributions associated with a geometric phase.
Furthermore, we compute the macroscopic static dielectric tensor of bulk \MoTe\ using density functional perturbation theory as described in Refs.\,\onlinecite{Baroni1986,Gajdos2006}, following Ref.\,\onlinecite{Laturia2018}.

\subsection{DFT based model Hamiltonian}

\begin{figure}[b!]
\includegraphics[width =0.41\textwidth]{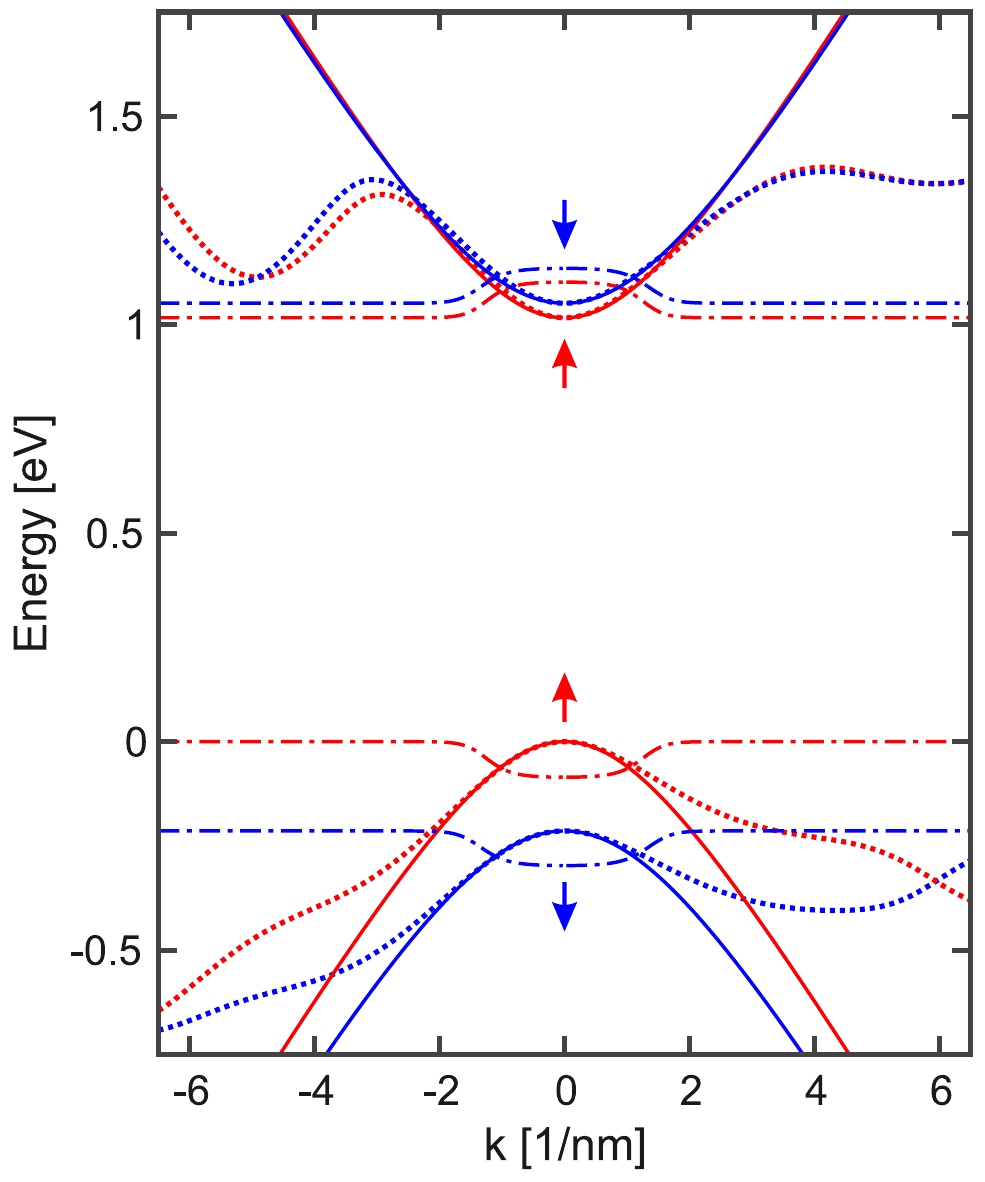}
\caption{\label{fig:02} Comparison of the relevant DFT bands (dotted) with the approximated unrenormalized relativistic band dispersion (solid).  Arrows indicate the spin of the correspondingly colored bands. The dashed-dotted lines show the equilibrium carrier distributions ($2.5$\,ps after excitation) for the excitation conditions discussed in the text. For the distributions the baseline is taken to be the edge of the corresponding band and the maximum values are set to be the corresponding chemical potentials.}
\end{figure}

To model the DFT band structure presented in the previous section, we include the two spin-split valence and conduction bands to obtain an effective four band Hamiltonian. As the different valleys are separated by large barriers, inter-valley scattering is expected to be significantly slower than intra-valley scattering and, on the ultrashort time scale, the valley index can be considered to be approximately conserved. Hence, we write the single particle part of the Hamiltonian as 
\begin{equation*}
H_0=\sum_{\alpha\bk}\epsilon_{\alpha\bk}c^\dagger_{\alpha\bk}c_{\alpha\bk},
\end{equation*}
where $\alpha$ combines the spin, valley, and band index. Using the $\bp\cdot\bA$ gauge, the light matter interaction is given by
\begin{equation*}
H_{\rm LM}=\frac{e}{m_0 c}\sum_{\alpha\alpha'\bk}\bA\cdot\bp_{\alpha\alpha'\bk}c^\dagger_{\alpha\bk}c_{\alpha'\bk},
\end{equation*}
where the interband momentum matrix elements are related to the DFT dipole matrix elements via $\frac{e\hbar}{m_0}\bp_{\alpha\alpha'\bk}=
(\epsilon_{\alpha'\bk}-\epsilon_{\alpha\bk})\bm{d}_{\alpha\alpha'}$. Whereas the side valleys at $\Sigma/\Lambda$ are modeled within the effective mass approximation, we treat the $K$ and $K'$ valleys utilizing the widely used massive Dirac-Fermion (MDF) model Hamiltonian\,\cite{xiao2012} to account for the geometric phase contained in the dipole matrix elements. The MDF Hamiltonian has the relativistic dispersion
\begin{equation}
 \epsilon^{c/v}_{i\mathbf{k}} = E_{F,i}\pm\frac{1}{2}\sqrt{\Delta_{i}^2+(2\hbar v_{F,i} k)^2},
\end{equation}
where $i=s\tau$ combines the spin and valley index, $\Delta_{i}$, $v_{F,i}$ and $E_{F,i}$ are the spin and valley dependent gap, Fermi velocity and Fermi level, respectively. 
Whereas the spin and valley-dependent band gaps are directly taken from our DFT calculations, the Fermi-velocities of the $\romA$- and $\romB$- bands are determined to reproduce the DFT band structure around the $K/K'$ points and listed in Table\,\ref{tab:MDFparameters}.  Within the MDF model, the non-vanishing dipole moments at the Dirac points are solely associated with the geometric phase or pseudo-spin. They are related to the Fermi velocity via $d^\pm_i=e\sqrt{2}\hbar v_{F,i}/\Delta_{i}$ and agree within less than $5\%$ with the DFT dipole matrix elements. The approximated band structure is shown together with the DFT bands and the equilibrium carrier distributions at the delay time $\tau=2.5$\,ps in Fig.\,\ref{fig:02}.

\begin{table}[b!]
\caption{Resulting MDF model parameters for \MoTe .}
\begin{ruledtabular}
\begin{tabular}{cccc}
 Band&$\Delta$\,[eV]&$\hbar v_F$\,[eV\AA]&$E_F$\,[eV]\\ \hline
 $\romA$&1.017&2.526&0.509    \\
 $\romB$&1.266&2.574&0.419   
\end{tabular}
\end{ruledtabular}
\label{tab:MDFparameters}
\end{table}

The Coulomb interaction Hamiltonian
\begin{eqnarray*}
H_C&=&\frac{1}{2}\sum_{\bq\neq 0} \sum_{\alpha\alpha'\beta\beta'\bk\bk'}V^{\alpha\beta\beta'\alpha'}_{\bq;\bk';\bk}
c^\dagger_{\alpha\bk-\bq}c^\dagger_{\beta\bk'+\bq}c_{\beta'\bk'}c_{\alpha'\bk}
\end{eqnarray*}
contains the quasi-2D Coulomb matrix elements
\begin{eqnarray*}
V^{\alpha\beta\beta'\alpha'}_{\bq;\bk';\bk}=\int_{ec}{\rm d}^3 r\int_{ec} & {\rm d}^3 r' & \, u^*_{\alpha\bk-\bq}(\br)u^*_{\beta\bk'+\bq}(\br')\notag\\ & \times & V_\bq(z-z') u_{\beta'\bk'}(\br')u_{\alpha'\bk}(\br)
\end{eqnarray*}
that are computed using the DFT wave functions. The Coulomb interaction potential $V_{\bq}(z-z')$ for the unexcited ML is determined from Poisson's equation according to Ref.\,\onlinecite{meckbach2018a}. {Here, we use the parameters $\epspar$ and $\epsz$ for the in- and out-of-plane dielectric constants based on bulk DFT calculations of \MoTe{}. From the previously stated bulk in-plane dielectric constant $\epspar^B$ we obtained the non-resonant 2D contribution $\epspar=15.32$ as described in Ref.\,\onlinecite{meckbach2018a}.} The so determined {'bare'} Coulomb interaction potential contains screening contributions from the dielectric environment and all remote bands, as well as the ground-state screening contributions from the valence and conduction bands. 

The interaction with longitudinal optical (LO) phonons, which has been shown to be the most effective  phonon coupling contribution {in monolayer \MoTe{}}\,\cite{sohier2016}, is contained in the Fr\"ohlich Hamiltonian
\begin{equation*}
H_{\rm e-LO}=\sum_{\alpha\bk,\bq}g^0_\bq c^{\dagger}_{\alpha,\bk+\bq}c_{\alpha,\bk}\left( b_{\bq}+b^{\dagger}_{-\bq}\right).
\end{equation*}
For the {'bare' Fr\"ohlich-}interaction matrix element $g^{0}_{\mathbf{q}}$, we use the explicit expression based on the analytical model of Sohier {\it et al.}\,\cite{sohier2016} {that, similarly to the 'bare' Coulomb interaction, already contains background screening contributions from the remote bands and dielectric environment.}

\subsection{Dirac-Bloch Equations}
To evaluate the material response after optical excitation, we compute the microscopic interband polarizations $P_{i\bk}=\langle c^\dagger_{i\romv\bk}c_{i \romc\bk}\rangle$
and occupation probabilities $ f^{\lambda}_{i\bk}=\langle c^\dagger_{i\lambda\bk}c_{i\lambda\bk}\rangle$ $(\lambda=\romc,\romv)$ from the Dirac Bloch equations (DBE) 
\begin{eqnarray}
\mathrm{i}\hbar\frac{d}{dt} P_{i\mathbf{k}} &=& (\Sigma^{\romc}_{i\mathbf{k}}-\Sigma^{\romv}_{i\mathbf{k}})P_{i\mathbf{k}}
	+(f^{\romv}_{i\mathbf{k}}-f^{\romc}_{i\mathbf{k}})\Omega_{i\mathbf{k}}\notag\\  
	&+& \left. \mathrm{i}\hbar\frac{d}{dt} P_{i\mathbf{k}}\right\vert_{\mathrm{corr.}},\label{eq:DBE_P}\\
\mathrm{i}\hbar\frac{d}{dt}f^{\romc/\romv}_{i\mathbf{k}}&=&
	\pm 2\mathrm{i}\Im\left[P_{i\mathbf{k}}\Omega^{*}_{i\mathbf{k}}\right]
	+\left. \mathrm{i}\hbar\frac{d}{dt} f^{\romc/\romv}_{i\mathbf{k}}\right\vert_{\mathrm{corr.}}.\label{eq:DBE_fc}
\end{eqnarray}
Here,
\begin{eqnarray}
\Sigma^{\romc}_{i\mathbf{k}} &=& \epsilon_{i\romc\mathbf{k}}-\sum_{\mathbf{k'}}
	\left[V^{\romc\romc\romc\romc}_{\mathbf{k-k'};\mathbf{k'};\mathbf{k}}-V^{\romc\romv\romc\romv}_{\mathbf{k-k'};\mathbf{k'};\mathbf{k}}\right]
	f^{\romc}_{i\mathbf{k'}}
\notag\\
&+& 
\sum_{\mathbf{k'}}\left[V^{\romc\romv\romc\romc}_{\mathbf{k-k'};\mathbf{k'};\mathbf{k}}P_{i\mathbf{k'}} + c.c.\right],
\label{eq:Renormalized_Ec}\\
\Sigma^{\romv}_{i\mathbf{k}} &=& \epsilon_{i\romv\mathbf{k}}-\sum_{\mathbf{k'}}
	\left[V^{\romv\romv\romv\romv}_{\mathbf{k-k'};\mathbf{k'};\mathbf{k}}-V^{\romv\romc\romv\romc}_{\mathbf{k-k'},\mathbf{k'},\mathbf{k}}\right]
	f^{\romv}_{i\mathbf{k'}}
\notag\\
&+&
 \sum_{\mathbf{k'}}\left[V^{\romv\romv\romv\romc}_{\mathbf{k-k'};\mathbf{k'};\mathbf{k}}P_{i\mathbf{k'}} + c.c.\right],
\label{eq:Renormalized_Ev}\\
\Omega_{i\mathbf{k}} &=&\frac{e}{m_0c}\bA\cdot\bp_{i\romc\romv\bk}
- 
\sum_{\mathbf{k'}}V^{\romc\romv\romv\romv}_{\mathbf{k-k'};\mathbf{k'};\mathbf{k}}
	\left(f^{\romv}_{i\mathbf{k'}}-f^{\romc}_{i\mathbf{k'}}\right)
\notag\\
&-& 
\sum_{\mathbf{k'}}\left[V^{\romc\romv\romv\romc}_{\mathbf{k-k'};\mathbf{k'};\mathbf{k}}P_{i\mathbf{k'}}
	+V^{\romc\romc\romv\romv}_{\mathbf{k-k'};\mathbf{k'};\mathbf{k}}P^{*}_{i\mathbf{k'}}\right]
\label{eq:Renormalized_Omega}
\end{eqnarray}
contain the Hartree-Fock contributions to the single particle energy renormalizations $\Sigma^{\romc/\romv}_{s\bk}$  and the renormalization to the Rabi energy $\Omega_{s\bk}$,  while all many-body correlations that arise from the two-particle Coulomb interaction and carrier-phonon scattering are contained within $\left.\frac{d}{dt} P_{s\mathbf{k}}\right\vert_{\mathrm{corr.}}$ and $\left.\frac{d}{dt} f^{\romc/\romv}_{s\mathbf{k}}\right\vert_{\mathrm{corr.}}$, respectively.

The DBE are formally equivalent to the semiconductor Bloch equations (SBE), but the expression for the renormalized Rabi and single particle energies differ by the Coulomb matrix elements of the type $V^{\romc\romv\romv\romv}_{\mathbf{k-k'};\mathbf{k'};\mathbf{k}}$ and $V^{\romc\romc\romv\romv}_{\mathbf{k-k'};\mathbf{k'};\mathbf{k}}$ referring to Auger- and pair creation/annihilation processes, where at least one of the particles changes its band index. Within the MDF model Hamiltonian, these 
Coulomb matrix elements contain the geometric phase and induce a coupling between a dark  static  interband polarization and the carrier populations. Consequently, the initial condition $f^\romc_{i\bk}=1-f^\romv_{i\bk}=P_{i\bk}=0$ does not correspond to a stationary solution of the DBE in the absence of an external field and hence, does not specify the system's ground state.

To determine the ground-state band renormalization, which is the initial state before the pump pulse arrives, we require a stationary solution of the Dirac-Bloch equations in the absence of an external field as described in Refs.\,\onlinecite{stroucken2011,stroucken2017,meckbach2018a}. This leads to the gap equations
\begin{eqnarray}
\label{gap1}
\tilde{\Delta}_{i\mathbf{k}} &=& \Delta_{i}+\frac{1}{2}\sum_{\mathbf{k'}}\,V_{\mathbf{|k-k'|}}\,		
	\frac{\tilde{\Delta}_{i\mathbf{k'}}}{\tilde{\epsilon}_{i\mathbf{k'}}},
\\
\tilde{v}_{i\mathbf{k}} &=& v_{F,i} + \frac{1}{2}\sum_{\mathbf{k'}}\,V_{\mathbf{|k-k'|}}\,
	\frac{k'}{k}\frac{\tilde{v}_{i\mathbf{k'}}}{\tilde{\epsilon}_{i\mathbf{k'}}} 
	\cos(\theta_{\mathbf{k}}-\theta_{\mathbf{k'}}),
	\label{Eq:Gap}
\label{gap2}
\end{eqnarray}
from which the ground-state quasi-particle dispersion is obtained via $\tilde{\epsilon}_{i\mathbf{k}}^{e/h}=\frac{1}{2}\sqrt{\tilde{\Delta}_{i\mathbf{k}}^2+\left(2\hbar\tilde{v}_{i\mathbf{k}} k\right)^2}$. {Similar to the Coulomb matrix elements $V^{\alpha\beta\beta'\alpha'}_{\bq;\bk';\bk}$, the 'bare' quasi-2D Coulomb interaction entering Eqs.\,(\ref{gap1})\,and\,(\ref{gap2}) already contains the previously mentioned background and ground-state screening contributions.} The solution of the gap equations yields a rigid shift of the single-particle dispersion with an interacting gap $\tilde{\Delta}_{i\mathbf{k}}$ that depends on the dielectric environment. For a suspended ML, we find a ground-state band renormalization of $549$\,meV for the A-band, that shrinks by $70$\,meV for a SiO$_{\mathrm{2}}$-supported \MoTe\ ML.

\begin{figure}[h!]
\includegraphics[width =0.45\textwidth]{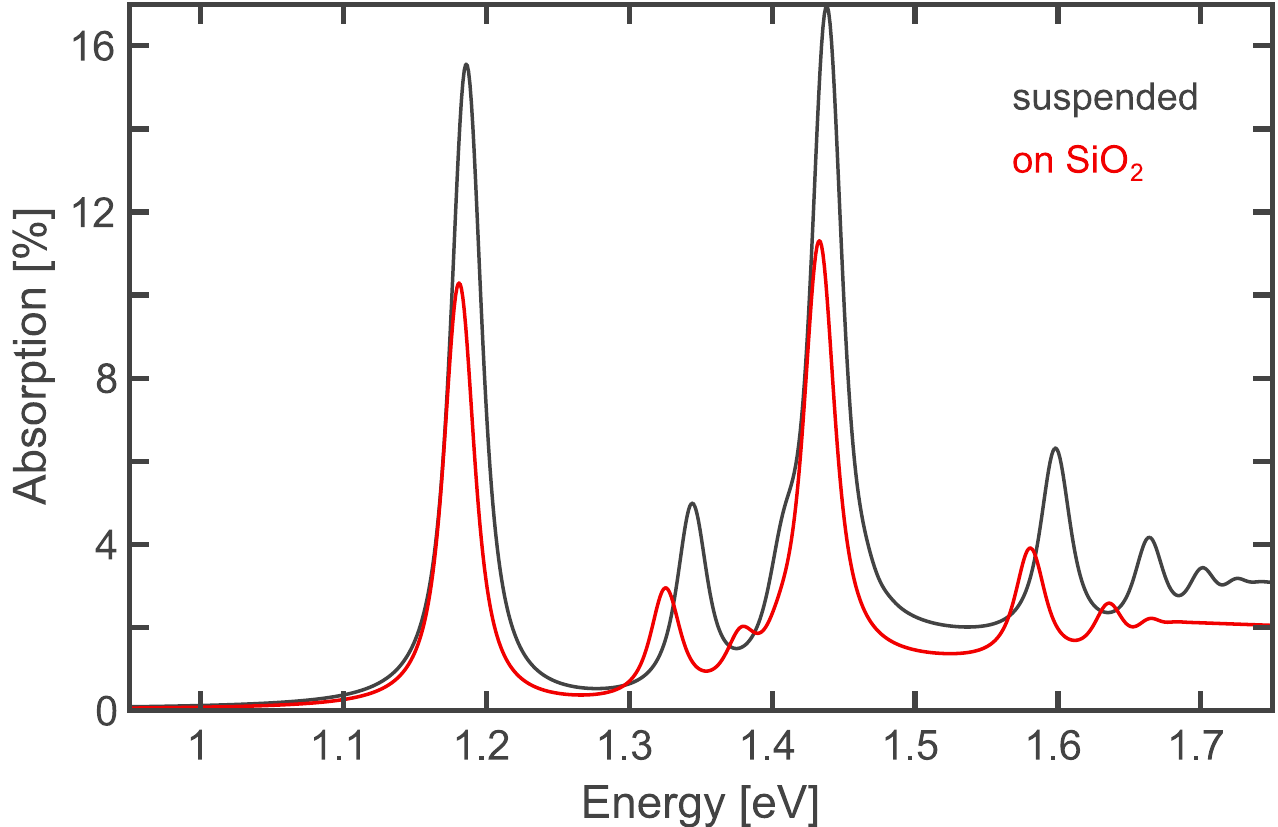}
\caption{\label{fig:03} Linear absorption spectra of a suspended (black) and SiO$_2$-supported (red) ML \MoTe. Here, we used  a phenomenological dephasing rate of $\hbar\gamma=10$\,meV. In order to match the experimentally observed data, the spectra have been shifted by $-0.044$\,eV.}
\end{figure}

In the low density limit, where correlation effects can be neglected, we find the transition energies for the lowest 1s-exciton resonance at $1.230$\,eV and $1.225$\,eV, for the suspended and SiO$_2$ supported \MoTe\ ML respectively, which is slightly above the  experimentally observed low-temperature resonance energy of about $1.18$\,eV for the SiO$_{\mathrm{2}}$-supported ML\,\cite{robert2016}. In order to allow for a direct comparison between the predicted excitation-induced modifications of the optical spectra with the experiment, we correct the bare DFT computed band gap and shift all spectra by $-0.044$\,eV. The resulting linear absorption spectra for a freely suspended and quartz-supported \MoTe\ ML are shown in Fig.\,\ref{fig:03} using a phenomenological dephasing rate of $\hbar\gamma=10$\,meV.

\subsection{Many-body correlations}\label{subsec:CoulombCorr}

Incoherent processes that lead to the dephasing of the microscopic polarizations and to the carrier relaxation dynamics are contained within $\left.\frac{d}{dt} P_{i\mathbf{k}}\right\vert_{\rm corr.}$ and $\left.\frac{d}{dt} f^{\romc/\romv}_{i\mathbf{k}}\right\vert_{\rm corr.}$, respectively. Using the notation $P^{\alpha\alpha'}_\bk=\langle c^\dagger_{\alpha\bk}c_{\alpha'\bk}\rangle$, $f^\alpha_\bk=P^{\alpha\alpha}_\bk$ for the single-particle expectation values, the Coulomb interaction leads to a contribution 

\begin{equation}
\left. i\hbar\frac{\rm d}{{\rm d}t}P^{\alpha\alpha'}_\bk\right|_{\mathrm{corr.}}^{\mathrm{el.}}=\sum_{\bq\neq 0}\left[\Delta I^{\alpha\alpha'}_{\bq;\bk}-\left(\Delta I^{\alpha'\alpha}_{\bq;\bk}\right)^*
\right],\label{eq:eesc_epsc}
\end{equation}
where
\begin{equation*}
 I^{\alpha\alpha'}_{\bq;\bk}=\sum_{\beta\gamma\gamma'}\sum_{\bk'}V^{\alpha'\gamma\gamma'\beta}_{-\bq;\bk';\bk-\bq}
C^{\alpha\gamma\gamma'\beta}_{\bq;\bk';\bk}
\end{equation*}
are the density-assisted transition/occupation probabilities that contain the two-particle expectation values
$C^{\alpha\gamma\gamma'\beta}_{\bq;\bk';\bk}=
\langle c^\dagger_{\alpha\bk}c^\dagger_{\gamma\bk'-\bq}c_{\gamma'\bk'}c_{\beta\bk-\bq}\rangle$.
In general, the two-particle expectation values can be divided into a Hartree-Fock (singlet) part and a correlated part according to
\begin{equation*}
C^{\alpha\gamma\gamma'\beta}_{\bq;\bk';\bk}=\left. C^{\alpha\gamma\gamma'\beta}_{\bq;\bk';\bk}\right|_S+\Delta C^{\alpha\gamma\gamma'\beta}_{\bq;\bk';\bk}
\end{equation*}
and it is easily verified that the Hartree-Fock contributions to the renormalized single particle energies and Rabi-energy given in Eqs.\,(\ref{eq:Renormalized_Ec})\,-\,(\ref{eq:Renormalized_Omega}) correspond to the singlet part 
$\left. I^{\alpha\alpha'}_{\bq;\bk}\right|_S$.

Physically, $ I^{\alpha\alpha'}_{\bq;\bk}$ describes Coulomb mediated transitions from all initial states $\beta\bk-\bq$ to the final state $\alpha\bk$ via the intermediate states 
${\gamma\bk'-\bq,\gamma'\bk'}$ with the Coulomb matrix elements ${V^{\alpha'\gamma\gamma'\beta}_{-\bq; \bk;\bk-\bq}}$. As mentioned above, the Coulomb matrix elements with $\beta\neq\alpha'$ and/or $\gamma\neq\gamma'$ correspond to Auger recombinations and pair creation/annihilation processes. Whereas these processes give a small contribution to the static ground-state renormalization of the valence bands contained in the singlet parts, they are strongly non resonant for the optically induced density dependent modifications contained in the correlated part. As verified numerically, the time scales for  Auger recombinations and pair creation/annihilation processes are four to five orders of magnitude longer than those of intraband scattering processes and will be neglected in the following. Additionally 
neglecting the weak orbital and $\bk,\bk'$-dependence of the Coulomb matrix elements, the expression for the density-assisted transition/occupation probabilities simplifies to
\begin{equation*}
 I^{\alpha\alpha'}_{\bq;\bk}=\tilde V_{\bq}\langle c^\dagger_{\alpha\bk}\rho_\bq c_{\alpha'\bk-\bq}\rangle,
\end{equation*}
where $\tilde V_\bq$ is {a quasi-2D Coulomb matrix element that again contains ground-state and background screening contributions only} and $\rho_\bq=\sum_{\beta\bk'}c^\dagger_{\beta\bk'-\bq}c_{\beta\bk'}$ is the density operator.

To derive an approximation for the correlated part of the  density assisted transition amplitudes, we employ a second-order cluster expansion where we derive the EOM for the relevant two-particle correlations and factorize the occurring three-particle expectation values into singlet and doublet contributions\,\cite{kira2006}. Within this approximation, the singlet factorizations act as source terms for the two-particle correlations, whereas the doublet contributions lead to a renormalization of the single particle energies, excitonic correlations, biexcitons and screening of the Coulomb interaction in the Hartree-Fock contributions. Assuming screening to be the dominant correlation effect at elevated densities, we 
write the EOM for the relevant two-particle correlations as
\begin{eqnarray}
i\hbar\frac{\rm d}{{\rm d}t}
\Delta C^{\alpha\beta\beta\alpha'}_{\bq;\bk';\bk}
&=&
\left( \Delta\Sigma^{\alpha\beta\beta\alpha'}_{\bq;\bk';\bk}-i\hbar\gamma_T\right)
\Delta C^{\alpha\beta\beta\alpha'}_{\bq;\bk';\bk}
\nonumber\\
&+&\left(f^{\beta}_{\bk'-\bq}-f^{\beta}_{\bk'}\right)I^{\alpha\alpha'}_{\bq,\bk}
+
S^{\alpha\beta\beta\alpha'}_{\bq;\bk';\bk}
\nonumber\\
&+&{\rm remaining\,\, doublets},
\label{eom_doublets_corr}
\end{eqnarray}
where we explicitly quoted only the doublet correlations that lead to the build-up of screening, $\Delta\Sigma^{\alpha\beta\beta\alpha'}_{\bq;\bk';\bk}=\Sigma_{\alpha'\bk-\bq} + \Sigma_{\beta\bk'}- \Sigma_{\beta\bk'-\bq}- \Sigma_{\alpha\bk}$, and included a phenomenological dephasing of the triplets $\gamma_T$. Note that in Eq.\,(\ref{eom_doublets_corr}), the density assisted transition probabilities $I^{\alpha\alpha'}_{\bq,\bk}$ contain both the singlet and correlated part and the remaining singlet sources are contained in $ S^{\alpha\beta\beta\alpha'}_{\bq;\bk';\bk}$. Using the shorthand notation $\bar f^\beta_\bk=1-f^\beta_\bk$, the singlet sources are explicitly given by

\begin{eqnarray}
 S^{\alpha\beta\beta\alpha'}_{\bq;\bk';\bk}&=&
\tilde V_\bq \left(
P^{\alpha\alpha'}_\bk f^\beta_{\bk'-\bq}\bar f^\beta_{\bk'}-P^{\alpha\alpha'}_{\bk-\bq} \bar f^\beta_{\bk'-\bq} f^\beta_{\bk'}
\right)
\nonumber\\
&+&\tilde V_{\bk-\bk'}P^{\alpha\beta}_\bk\sum_\gamma
P^{\beta\gamma}_{\bk'-\bq}\left(P^{\gamma\alpha'}_{\bk-\bq}-\delta_{\gamma\alpha'}\right)
\nonumber\\
&-&\tilde V_{\bk-\bk'}P^{\alpha\beta}_{\bk'}\sum_\gamma P^{\gamma\alpha'}_{\bk-\bq}\left(P^{\beta\gamma}_{\bk'-\bq}-\delta_{\gamma\beta}\right)
\nonumber\\
&+&
\tilde V_{\bk-\bk'}\left(P^{\beta\alpha'}_{\bk'-\bq}-P^{\beta\alpha'}_{\bk-\bq}\right)\sum_\gamma P^{\alpha\gamma}_\bk P^{\gamma\beta}_{\bk'}
\label{singletsources}
\end{eqnarray}
and treated on the level of a second Born approximation. 

\begin{figure}[b!]
\includegraphics[width =0.49\textwidth]{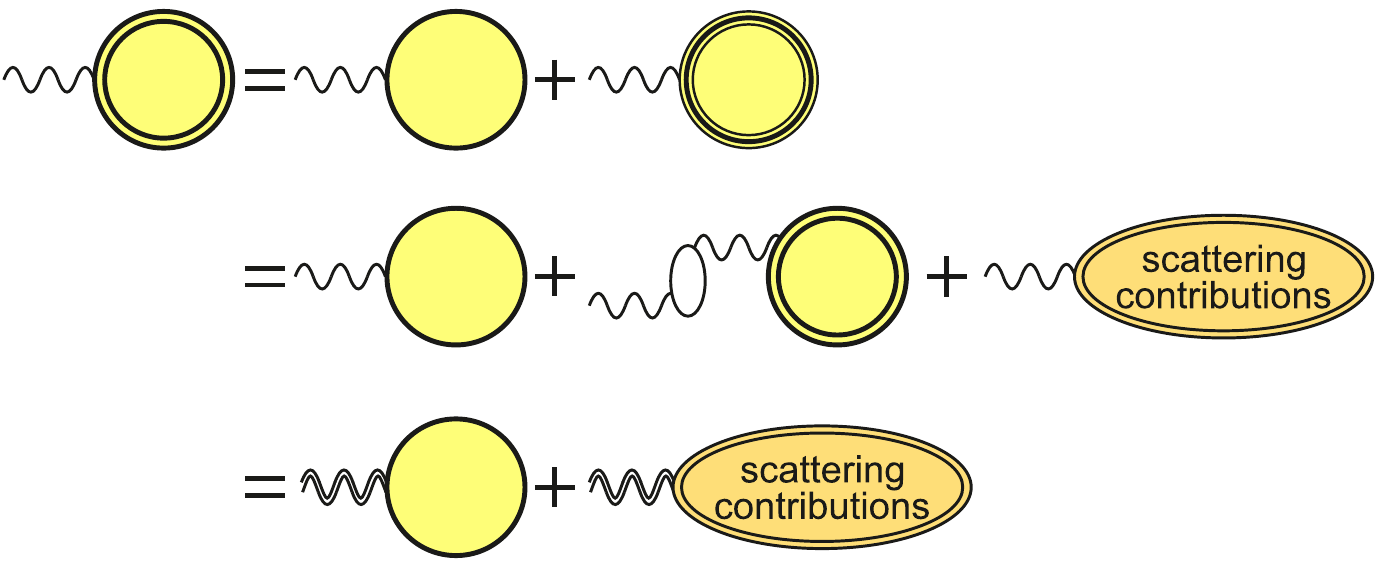}
\caption{
Diagrammatic representation of the density assisted transition/occupation probabilities. The full density assisted transition amplitudes are represented by twofold contoured circles whereas the singlet and correlated parts are represented by simple and threefold contoured circles, respectively. Similarly, simple and doubly contoured wiggles represent the 'bare' and screened Coulomb interaction, respectively. The upper line corresponds to the division into the singlet (Hartree-Fock) and correlated part and the second and third lines represent the first and second line of Eq.\,(\ref{eq:I1}).
\label{diagram}}
\end{figure}

If one assumes quasi-static single-particle distributions $f^\beta_\bk$ and neglects the remaining doublet contributions, one can analytically solve Eq.\,(\ref{eom_doublets_corr}) in frequency space\,\cite{kira2006}. A subsequent summation over $\beta$ and $\bk'$ yields the closed expression
\begin{eqnarray}
 I^{\alpha\alpha'}_{\bq;\bk}(\omega)&=&
 \left.  I^{\alpha\alpha'}_{\bq;\bk}(\omega)\right|_S
+
\tilde V_\bq\Pi^{\alpha\alpha'}_{\bq;\bk}(\omega+i\gamma_T)  I^{\alpha\alpha'}_{\bq,\bk}(\omega)
\nonumber\\
&+&\tilde V_\bq T^{\alpha\alpha'}_{\bq;\bk}(\omega)\nonumber\\
&=& W^{\alpha\alpha'}_{\bq;\bk} (\omega+i\gamma_T)\sum_{\beta\bk'}\left. C^{\alpha\beta\beta\alpha'}_{\bq;\bk';\bk}(\omega)\right|_S\nonumber\\
&+& W^{\alpha\alpha'}_{\bq;\bk} (\omega+i\gamma_T)T^{\alpha\alpha'}_{\bq;\bk}(\omega),\label{eq:I1}
\end{eqnarray}
where the screened Coulomb matrix element $W^{\alpha\alpha'}_{\bq;\bk}$ is given by Dyson's equation 
{
\begin{equation*}
W^{\alpha\alpha'}_{\bq;\bk}(\omega) = \tilde{V}_\bq+\tilde{V}_\bq
\Pi^{\alpha\alpha'}_{\bq;\bk}(\omega)W^{\alpha\alpha'}_{\bq;\bk}(\omega)
\end{equation*}
}
and
\begin{eqnarray*}
\Pi^{\alpha\alpha'}_{\bq;\bk}(\omega)&=&\Pi_\bq(\omega+ (\Sigma_{\alpha\bk}-\Sigma_{\alpha'\bk-\bq})/\hbar),\\
\Pi_\bq(\omega)&=&\sum_{\beta\bk'}\frac{f^\beta_{\bk'-\bq}-f^\beta_{\bk'}}{\hbar\omega+\Sigma_{\beta\bk'-\bq}-\Sigma_{\beta\bk'}}
\end{eqnarray*}
is the standard Lindhard polarization function. Furthermore, we define
\begin{equation*}
T^{\alpha\alpha'}_{\bq;\bk}(\omega)=\sum_{\beta\bk'}\frac{S^{\alpha\beta\beta\alpha'}_{\bq;\bk';\bk}(\omega)}{\hbar\omega-\Delta\Sigma^{\alpha\beta\beta\alpha'}_{\bq;\bk';\bk}+i\hbar\gamma_T}.
\end{equation*}
Hence, the doublet contributions explicitly written in Eq.\,(\ref{eom_doublets_corr}) lead to screening of the  Coulomb potential both in the Hartree-Fock and scattering contributions, as shown schematically in Fig.\,\ref{diagram}. Note that at this level of approximation, the energy denominators in the scattering integrals and Lindhard polarization function still contain the unscreened Hartree-Fock renormalizations. However, it can be shown that screening in these contributions is introduced by the inclusion of the next level cluster expansion\,\cite{kira2006} and we will replace all unscreened energy renormalizations $\Sigma^\alpha_\bk$ by their screened counterparts $\tilde\Sigma^\alpha_\bk$ in the following.

If one would neglect the remaining scattering integrals contained in $\sum_\bq W^{\alpha\alpha'}_{\bq;\bk} T^{\alpha\alpha'}_{\bq;\bk}$, one would thus arrive at the level of the screened Hartree-Fock approximation where all the Coulomb matrix elements in Eqs.\,(\ref{eq:Renormalized_Ec})\,-\,(\ref{eq:Renormalized_Omega}) are replaced by their screened counterparts. The remaining scattering contributions predominantly lead to carrier relaxation and excitation induced dephasing of the coherent interband polarization. They contain only small additional renormalizations of the single-particle dispersion. Hence, the screened Hartree-Fock renormalizations of the single-particle energies provide a useful measure for the band-gap renormalization in the presence of excited carriers. In particular, in a fully incoherent quasi-static equilibrium situation, a steady state solution of the screened DBE equations yields the density dependent gap equations
\begin{eqnarray}
\label{gap1_dens}
\tilde{\Delta}_{i\mathbf{k}} = \Delta_{i}&+&\frac{1}{2}\sum_{\mathbf{k'}}\,W_{\mathbf{|k-k'|}}\,		
	\frac{\tilde{\Delta}_{i\mathbf{k'}}}{\tilde{\epsilon}_{i\mathbf{k'}}}
\left(f^\romv_{i\bk'}-f^\romc_{i\bk'}\right),
\\
\tilde{v}_{i\mathbf{k}} = v_{F,i} &+& \frac{1}{2}\sum_{\mathbf{k'}}\,W_{\mathbf{|k-k'|}}\,
	\frac{k'}{k}\frac{\tilde{v}_{i\mathbf{k'}}}{\tilde{\epsilon}_{i\mathbf{k'}}} 
	\cos(\theta_{\mathbf{k}}-\theta_{\mathbf{k'}})\nonumber\\
&\times& \left(f^\romv_{i\bk'}-f^\romc_{i\bk'}\right).
	\label{Eq:Gap_dens}
\label{gap2_dens}
\end{eqnarray}
where $W_\bq$ is the statically screened Coulomb interaction. 

Eq.\,(\ref{eq:I1}) together with Eq.\,(\ref{singletsources}) provides an efficient scheme to compute the Coulomb correlations numerically. As a consequence of the strong Coulomb interaction in TMDCs and the large associated exciton binding, it is important to properly include the full frequency dependence of the screened interband matrix elements $W^{\romv\romc}_{\bq;\bk}$ which enter the scattering contributions of the microscopic polarizations and lead to memory effects beyond the Markov approximation.

In addition to the Coulomb correlations, the inclusion of the Fr\"ohlich Hamiltonian introduces the scattering of excited charge carriers within their bands by the absorption or emission of LO phonons. These are evaluated on the level of quantum-kinetic theory in second Born approximation. Our analysis shows that, similarly as for the Coulombic scattering rates, it is sufficient to treat the phonon scattering rates of the quasi-static carrier distribution functions within the Markov approximation, whereas it is crucial to maintain the full frequency dependency for the polarization-phonon scattering rates. The resulting phononic contributions to the electron and polarization scattering rates are given by
\begin{widetext}
\begin{eqnarray}
\left. \frac{d}{dt} f^{\romc}_{s\mathbf{k}}\right\vert_{\mathrm{corr.}}^{\mathrm{ph.}}&=&
	 \frac{2\pi}{\hbar}\sum_{\mathbf{q}}g^{0}_{\mathbf{q}}g^{\romc\romc}_{\bq;\bk+\bq}\,
	\mathcal{D}_{\eta}\left(\tilde{\Sigma}^{\romc}_{s\mathbf{k+q}}-\tilde{\Sigma}^{\romc}_{s\mathbf{k}}-\hbar\omega_{\mathbf{q}}\right)
	\left[(n_{\mathbf{q}}+1)f^{\romc}_{s\mathbf{k+q}}\bar{f}^{\romc}_{s\mathbf{k}}-
	n_{\mathbf{q}}f^{\romc}_{s\mathbf{k}}\bar{f}^{\romc}_{s\mathbf{k+q}}\right]\notag\\
&+&
	\frac{2\pi}{\hbar}\sum_{\mathbf{q}}g^{0}_{\mathbf{q}}g^{\romc\romc}_{\bq;\bk}\,
	\mathcal{D}_{\eta}\left(\tilde{\Sigma}^{\romc}_{s\mathbf{k-q}}-\tilde{\Sigma}^{\romc}_{s\mathbf{k}}+\hbar\omega_{\mathbf{q}}\right)
	\left[n_{\mathbf{q}}f^{\romc}_{s\mathbf{k-q}}\bar{f}^{\romc}_{s\mathbf{k}}-
	(1+n_{\mathbf{q}})f^{\romc}_{s\mathbf{k}}\bar{f}^{\romc}_{s\mathbf{k-q}}\right]
\label{eq:phsc}\\
\mathrm{i}\hbar\left.\frac{d}{dt}P_{s\mathbf{k}}\right\vert_{\mathrm{corr.}}^{\mathrm{ph.}} 
&=& \mathrm{i}\hbar\left.\frac{d}{dt}P_{s\mathbf{k}}\right\vert_{\mathrm{corr.}}^{\romc,\mathrm{ph.}}
+ \mathrm{i}\hbar\left.\frac{d}{dt}P_{s\mathbf{k}}\right\vert_{\mathrm{corr.}}^{\romv,\mathrm{ph.}},
\label{eq:ppsc1}\\
\mathrm{i}\hbar\left.\frac{d}{dt} P_{s\mathbf{k}}\right\vert_{\mathrm{corr.}}^{\romc,\mathrm{ph.}}
&=&
\mathcal{F}\left[
	\sum_{\mathbf{q}}\,g^{0}_{\mathbf{q}}g^{\romv\romc}_{\mathbf{q};\mathbf{k}}\,\left\{
	\frac{\bar{f}^{\romc}_{s\mathbf{k-q}}n_{\mathbf{q}}+f^{\romc}_{s\mathbf{k-q}}(1+n_{\mathbf{q}})}
	{\hbar(\omega+\omega_{\mathbf{q}})+\tilde{\Sigma}^{\romv}_{s\mathbf{k}}-\tilde{\Sigma}^{\romc}_{s\mathbf{k-q}}+\mathrm{i}\eta}
	+\frac{\bar{f}^{\romc}_{s\mathbf{k-q}}(1+n_{\mathbf{q}})+f^{\romc}_{s\mathbf{k-q}}n_{\mathbf{q}}}
	{\hbar(\omega-\omega_{\mathbf{q}})+\tilde{\Sigma}^{\romv}_{s\mathbf{k}}-\tilde{\Sigma}^{\romc}_{s\mathbf{k-q}}+\mathrm{i}\eta}
	\right\}P_{s\mathbf{k}}\right.\notag\\
&-&
	\left.\sum_{\mathbf{q}}\,g^{0}_{\mathbf{q}}g^{\romv\romc}_{\mathbf{q};\mathbf{k+q}}\,\left\{
	\frac{\bar{f}^{\romc}_{s\mathbf{k}}n_{\mathbf{q}}+f^{\romc}_{s\mathbf{k}}(1+n_{\mathbf{q}})}
	{\hbar(\omega+\omega_{\mathbf{q}})+\tilde{\Sigma}^{\romv}_{s\mathbf{k+q}}-\tilde{\Sigma}^{\romc}_{s\mathbf{k}}+\mathrm{i}\eta}
	+\frac{\bar{f}^{\romc}_{s\mathbf{k}}(1+n_{\mathbf{q}})+f^{\romc}_{s\mathbf{k}}n_{\mathbf{q}}}
	{\hbar(\omega-\omega_{\mathbf{q}})+\tilde{\Sigma}^{\romv}_{s\mathbf{k+q}}-\tilde{\Sigma}^{\romc}_{s\mathbf{k}}+\mathrm{i}\eta}
	\right\}P_{s\mathbf{k+q}}\right],
\label{eq:ppsc2}\\
\mathrm{i}\hbar\left.\frac{d}{dt} P_{s\mathbf{k}}\right\vert_{\mathrm{corr.}}^{\romv,\mathrm{ph.}}
&=&
\mathcal{F}\left[
	\sum_{\mathbf{q}}\,g^{0}_{\mathbf{q}}g^{\romv\romc}_{\mathbf{q};\mathbf{k+q}}\,\left\{
	\frac{f^{\romv}_{s\mathbf{k+q}}n_{\mathbf{q}}+\bar{f}^{\romv}_{s\mathbf{k+q}}(1+n_{\mathbf{q}})}
	{\hbar(\omega+\omega_{\mathbf{q}})+\tilde{\Sigma}^{\romv}_{s\mathbf{k+q}}-\tilde{\Sigma}^{\romc}_{s\mathbf{k}}+\mathrm{i}\eta}
	+\frac{f^{\romv}_{s\mathbf{k+q}}(1+n_{\mathbf{q}})+\bar{f}^{\romv}_{s\mathbf{k+q}}n_{\mathbf{q}}}
	{\hbar(\omega-\omega_{\mathbf{q}})+\tilde{\Sigma}^{\romv}_{s\mathbf{k+q}}-\tilde{\Sigma}^{\romc}_{s\mathbf{k}}+\mathrm{i}\eta}
	\right\}P_{s\mathbf{k}}\right.\notag\\
&-&
	\left.\sum_{\mathbf{q}}\,g^{0}_{\mathbf{q}}g^{\romv\romc}_{\mathbf{q};\mathbf{k}}\,\left\{
	\frac{f^{\romv}_{s\mathbf{k}}n_{\mathbf{q}}+\bar{f}^{\romv}_{s\mathbf{k}}(1+n_{\mathbf{q}})}
	{\hbar(\omega+\omega_{\mathbf{q}})+\tilde{\Sigma}^{\romv}_{s\mathbf{k}}-\tilde{\Sigma}^{\romc}_{s\mathbf{k-q}}+\mathrm{i}\eta}
	+\frac{f^{\romv}_{s\mathbf{k}}(1+n_{\mathbf{q}})+\bar{f}^{\romv}_{s\mathbf{k}}n_{\mathbf{q}}}
	{\hbar(\omega-\omega_{\mathbf{q}})+\tilde{\Sigma}^{\romv}_{s\mathbf{k}}-\tilde{\Sigma}^{\romc}_{s\mathbf{k-q}}+\mathrm{i}\eta}
	\right\}P_{s\mathbf{k-q}}\right],
\label{eq:ppsc3}
\end{eqnarray}
\end{widetext}
and a similar equation holds for the valence-band distribution functions. {Here, $\pi\mathcal{D}_{\eta}(x)=\frac{\eta}{x^2+\eta^2}$ denotes the numerical energy-conserving function, $n_{\mathbf{q}}$ is the phonon occupation number, $\hbar\omega_{\mathbf{q}}=27.72 $\,meV\,\cite{sohier2016} is the corresponding LO-phonon energy, and $\mathcal{F}[f]$ denotes the Fourier transform of function $f$. 
The inclusion of the screened Fr{\"o}hlich interaction $g^{\alpha\alpha'}_{\mathbf{q},\mathbf{k}}(\omega) = g^{0}_{\mathbf{q}}+g^0_{\mathbf{q}}\Pi^{\alpha\alpha'}_{\mathbf{q},\mathbf{k}}(\omega)g^{\alpha\alpha'}_{\mathbf{q},\mathbf{k}}(\omega)$ accounts for screening contributions arising from the excited charge carriers in addition to the background screening contributions of the remote bands and dielectric environment already included in $g^0_{\mathbf{q}}$. Thus, Coulomb and phonon-coupling matrix elements are  treated on the same level of approximation.
} 

\section{Numerical results}

In this section, we present the results of our numerical analysis of the excitation dynamics and gain build-up in monolayer \MoTe . As physical conditions, we assume a pump-probe scenario, where we consider a room-temperature ML of \MoTe , which has been placed on a quartz substrate and is excited by a high-intensity linear-polarized optical pump pulse ($E_0\sim 1.25$\,MV/cm). The central pump frequency is chosen to be slightly above the interacting B-band gap and the pump pulse has a full width at half maximum (FWHM) of $333$\,fs corresponding to a photon density of $1.8\times 10^{15}$\,cm$^{-2}$ (a pump fluence of $520$\,$\mathrm{\mu}$J/cm$^{2}$). For these excitation conditions, it is ensured that virtually all carriers are created in the $K$ and $K'$ valleys.

For the pump simulations, we solve the Dirac-Bloch equations\,(\ref{eq:DBE_P})\,and\,(\ref{eq:DBE_fc}) in the time domain. The optically induced interband polarizations lead to the generation of excited charge carriers and the subsequent carrier relaxation dynamics is computed from the carrier-carrier and carrier-phonon scattering contributions, Eqs.\,(\ref{eq:eesc_epsc})\,-\,(\ref{eq:I1}) for $\alpha=\alpha'$ and Eq.\,(\ref{eq:phsc}), respectively. For the carrier dynamics, the numerically most important effect of the many-body correlations of the interband polarizations is the replacement of the 'bare' Coulomb potential by its screened counterpart in the Hartree-Fock contributions, whereas the detailed excitation induced dephasing of the  interband polarizations play only a minor role. As verified numerically, it is sufficient to compute the carrier dynamics within the Markov approximation and the excitation-induced band-gap renormalization is determined from the density dependent gap equations (\ref{gap1_dens}) and (\ref{gap2_dens}). {They yield the renormalized single-particle bands wherein the excited carriers relax on the level of screened Hartree-Fock approximation. We then define the excitation-induced band-gap renormalization as the density-dependent change of the gaps between the spin-split renormalized single-particle bands relative to the respective gaps in the low-density limit. }

For the probe pulse and for equilibrium configurations, we solve Eq.\,(\ref{eq:DBE_P}) in frequency domain via a matrix-inversion scheme, where the carrier distribution functions are quasi-static and  we restrict ourselves to the terms linear in $P_{s\mathbf{k}}$ in the singlet sources.  However, in order to predict the correct line shapes of the optical spectra, it turns out to be crucial to include the detailed and fully dynamical many-body correlations of the interband polarizations, Eqs.\,(\ref{eq:eesc_epsc})\,-\,(\ref{eq:I1}) for $\alpha/\alpha'=\romv/\romc$ and Eq.\,(\ref{eq:ppsc1})\,-\,(\ref{eq:ppsc3}).

\subsection{Carrier dynamics and band-gap renormalization}
\begin{figure}[h!]
\includegraphics[width =0.45\textwidth]{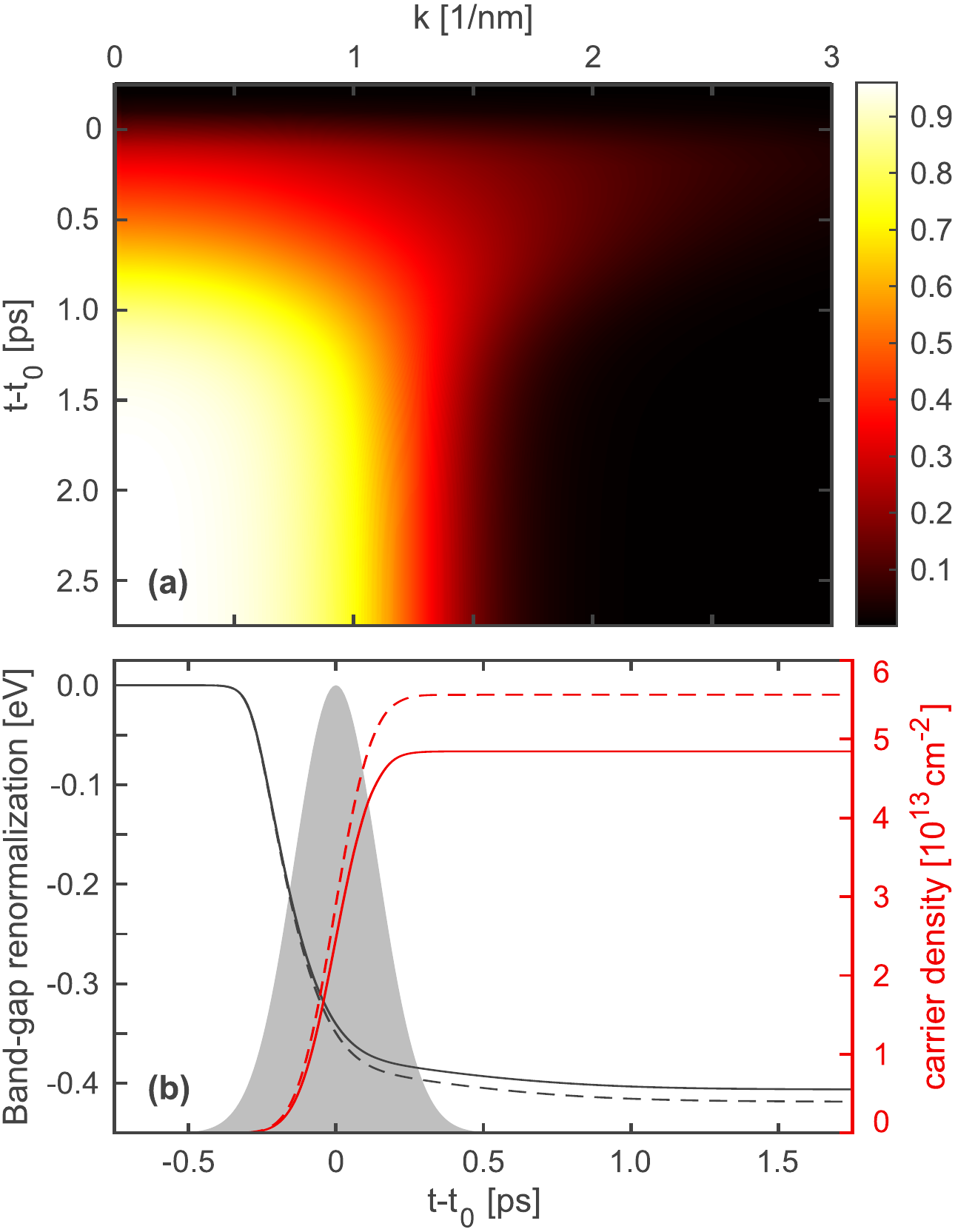}
\hspace{0.05\textwidth}
\caption{\label{fig:5} (a) Dynamics of the $\romA$-band electron distribution in the vicinity of the $K/K'$ points after excitation with a $333$\,fs pump pulse with peak amplitude of $1.3$\,MV/cm at $t=t_0$. (b) Time evolution of the excitation-induced band-gap renormalization (black) and the density of excited charge carriers (red). The solid lines correspond to the $\romA$-band properties, whereas the dashed lines show the $\romB$-band properties. The gray shaded area indicates the envelope of the optical pump pulse.}
\end{figure}

In Fig.\,\ref{fig:5}\,(a), we plot the excitation and subsequent relaxation dynamics of the $\romA$-band electron distribution function $f^{\rome}_{\romA,\mathbf{k}}(t)=f^{\romc}_{\romA,\mathbf{k}}(t)$ in the vicinity of the $K/K'$ points. Since conduction and valence bands with the same spin and valley indices have identical effective masses within the MDF model, the evolution of the hole distribution function $f^{\romh}_{\romA,\mathbf{k}}(t)=\bar{f}^{\romv}_{\romA,\mathbf{-k}}(t)$ is identical to that of the electrons. The excitation dynamics of the $\romB$-band distribution functions are similar and not shown here.

Due to the strong Coulomb interaction, carrier-carrier scattering is extremely efficient and drives the carrier distributions into hot quasi-equilibrium distributions within a few femtoseconds. Here, the carrier temperature reaches $T=2350$\,K $10$\,fs after the pump maximum has interacted with the sample. This ultrafast carrier-carrier scattering quickly redistributes the pump injected carriers to energies near the band gap and away from the excitation energy, thus almost completely preventing the accumulation of carriers at the excitation energy and effectively removing the associated Pauli-blocking of the absorption during the excitation process. The result is a highly efficient generation of excited charge carriers which is limited only by the absorption coefficient of the unexcited layer. After the pump pulse has passed, we find a total carrier density of $1.040\times 10^{14}$\,cm$^{-2}$, or equivalently, $5.20\times 10^{13}$\,cm$^{-2}$ generated electron hole-pairs, corresponding to an absorption of about $2.9$\,\% of the incoming photons. Due to the Coulomb enhancement of the above band-edge absorption, this value is slightly larger than the universal low-density continuum absorption of $\pi\alpha=2.3$\,\% for non-interacting Dirac Fermions.

The initial ultrafast relaxation into hot quasi-equilibrium distributions is followed by a phonon-induced thermalization that takes about $2.5$\,ps until a quasi-equilibrium at the temperature of the phonon bath ($300$\,K) is reached. At $\tau=0.1$\,ps and $\tau= 1$\,ps, we find intermediate temperatures of $T=2175$\,K and $T=555$\,K, respectively. This relaxation time is about twice as fast as in conventional semiconductors and based on the efficient phonon coupling in ML \MoTe\,\cite{sohier2016}. 

The time evolution of the excitation-induced band-gap renormalization (black) and the total carrier density with given spin and valley index $n_{i}(t)=\frac{1}{\mathcal{A}}\sum_{\alpha,\mathbf{k}} f^{\alpha}_{i\mathbf{k}}(t)$ (red) is shown in Fig.\,\ref{fig:5}\,(b). With solid lines, we depict the respective $\romA$-band properties, whereas the B-band properties are plotted using dashed lines. The gray shaded area shows the envelope of the Gaussian shaped optical pump pulse centered around $t_0$. Due to the initially nearly resonant excitation with the B-band gap, the final amount of charge carriers in the $\romB$-bands ($n_\romB=5.56\times 10^{13}$\,cm$^{-2}$) is slightly higher than in the $\romA$-bands ($n_\romA=4.84\times 10^{13}$\,cm$^{-2}$). Note that due to the opposite spin, there are no relaxation processes between the $\romB$- and the $\romA$-bands on the timescales of interest here. 

As can be recognized in Fig.\,\ref{fig:5}\,(b), the build-up of populations during the excitation process is accompanied by an almost instantaneous large shrinkage of the band gap, followed by a much slower further reduction. The band-edge shrinkage results from combined screening and phase-space-filling effects. As in conventional two-dimensional semiconductors, the screening wave number is proportional to the carrier occupation at $k=0$. Due to the ultrafast Coulomb scattering, the major contribution to screening develops within the time scale of the pump pulse with a correspondent reduction of the band gap of about $389$\,meV within the first $0.4$\,ps. Once the amount of excited charge carriers has saturated, the band edge can only be further reduced by phase-space filling. This leads to an additional reduction of about $17$\,meV on the time scale of the thermalization, yielding the total excitation-induced band-gap renormalizations of $406$\,meV ($\romA$-gap) and $419$\,meV ($\romB$-gap), respectively, for the investigated excitation conditions. These results are in good agreement with our previous findings in the equilibrium regime\,\cite{meckbach2018b} and reported experimental observations\,\cite{chernikov2015,ulstrup2016} on similar systems.

\subsection{Evolution of optical spectra and build-up of optical gain}

\begin{figure}[h!]
\includegraphics[width =0.45\textwidth]{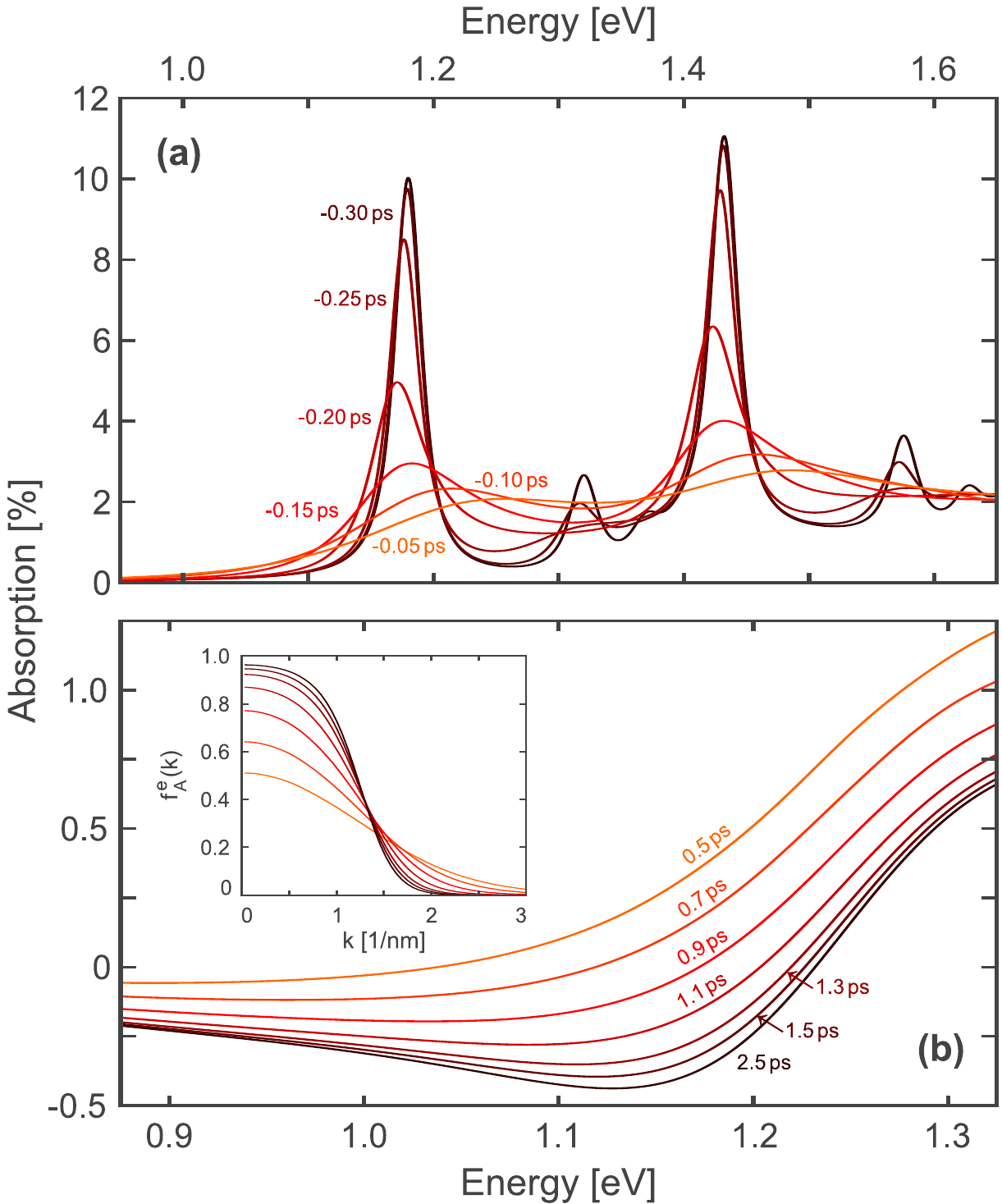}
\caption{\label{fig:6} Optical absorption spectra at distinct pump-probe delays. (a) Excitation regime. Pump-probe delays increase from $\tau=-0.30$\,ps to $\tau=-0.05$\,ps in $0.05$\,ps steps. The black solid line represents the low-density limit before the excitation. (b) Thermalization regime. Pump-probe delays increase from $\tau=0.5$\,ps to $\tau=1.5$\,ps in $0.2$\,ps steps. The absorption spectrum in the quasi-equilibrium limit ($\tau\geq2.5$\,ps) is shown in black. The corresponding $\romA$-band distribution functions are depicted in the inset.}
\end{figure}

In Fig.\,\ref{fig:6}, we present the time evolution of optical absorption/gain spectra computed as linear response to an ultra-short, low-intensity probe pulse for different delay times $\tau=t-t_0$. To cover the wide relevant energy range of several hundred meV, we choose a temporal width of $10$\,fs for the probe pulse. The use of such an ultrashort probe pulse also provides the necessary time resolution to study the evolution of the optical response during the excitation process, which is shown in Fig.\,\ref{fig:6}\,(a). Here, the pump-probe delay increases from $\tau=-0.30$\,ps to $\tau=-0.05$\,ps in $0.05$\,ps steps. For comparison, the linear absorption spectrum of the unexcited ML is depicted in black. We notice an initial increase of the excited carrier density leading to dephasing, excitation-induced band-edge shrinkage, and reduction of the exciton binding energy. As a consequence of compensating effects, we observe practically no shift of the exciton resonance position under the given excitation conditions. Excitation-induced dephasing increases the $1s$-exciton linewidth from $2.3$\,meV at a delay time $\tau=-0.25$\,ps to $42.2$\,meV at $\tau=-0.15$\,ps. Note, that we include an additional dephasing constant of $10$\,meV for the microscopic polarizations to assure convergence in the zero-density limit. At $\tau=-0.05$\,ps and a density of about $3.36\times 10^{13}$\,cm$^{-2}$ the exciton resonance is completely bleached out, marking the Mott-density.   

In Fig.\,\ref{fig:6}\,(b), we show the optical absorption in the thermalization regime. Here, pump-probe delays increase from $\tau=0.5$\,ps to $\tau=1.5$\,ps in $0.2$\,ps steps. The inset displays snapshots of the corresponding $\romA$-band electron distribution functions. In this time regime, no additional excited charge carriers are generated, but thermalization relaxes the existing carriers into quasi-equilibrium at the lattice temperature ($300$\,K). As a consequence of the high barriers between the $K/K'$ and $\Sigma/\Lambda$ valleys, no significant percentage of the excited carriers can reach the side valleys on the fast carrier relaxation time scale. Therefore, the total carrier densities in the $K/K'$ valleys is practically conserved. The absorption spectrum for quasi-equilibrium conditions is depicted in black in Fig.\,\ref{fig:6}\,(b). With decreasing temperature of the quasi-equilibrium distributions, the occupation probabilities near the band gap increase. About $0.5$\,ps after the pump pulse, inversion with $(1-f^{\rome}_{i\mathbf{k}}-f^{\romh}_{i\mathbf{k}})<0$ is reached and optical gain (negative absorption) appears in the spectrum. Inversion and gain increase with increasing cooling of the carriers. After about $2.5$\,ps thermal equilibrium is reached, where we observe broad A-band optical gain with a peak energy of $1.13$\,eV, slightly below the low density $\romA$-exciton resonance. {Note that while the inclusion of the carrier-phonon scattering is crucial for the thermalization of the carriers, linewidths and energy renormalizations within the optical spectra are dominated by polarization-carrier rather than polarization-phonon scattering contributions.} The maximum gain approaches a value of $0.5$\,\% amplification of the incoming light, which is clearly below the theoretical upper limit of $\pi\alpha/2=1.15$\,\% for the free carrier, single band case. Therefore, continuum absorption of the $\romA$-band overcompensates the gain of the $\romB$-band at higher energies yielding a net absorption in the frequency range of the $\romB$-exciton. 

\subsection{Influence of the side valleys\label{sec:equilibration}}

The full DFT band structure displays six side minima in the conduction band located at the $\Sigma/\Lambda$ points of the Brillioun zone. For the previously considered excitation conditions, carrier equilibration leads to a net electron drain from the $K/K'$ to the $\Sigma/\Lambda$ valleys as well as a hole drift between the $K$ and $K'$ valleys. To compute this side-valley drain and hole drift and its influence on the optical gain spectra, we consider an equilibrium situation where all electrons and holes have relaxed to a common chemical potentials. 
The side valleys are treated within the effective mass approximation and the influence of their populations is considered self-consistently including their contribution to screening and excitation induced renormalizations. 

\begin{figure}[htb]
\includegraphics[width =0.45\textwidth]{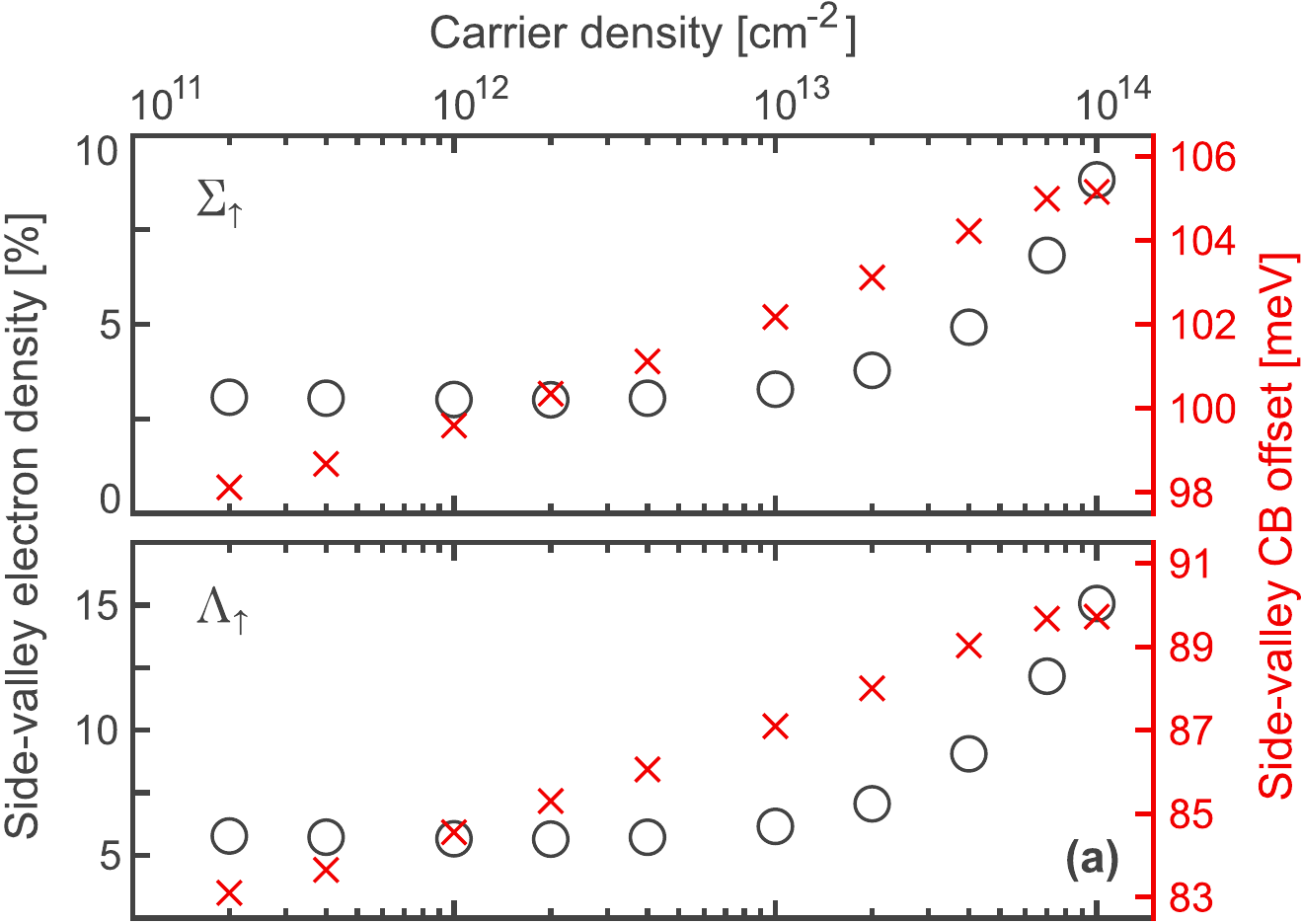}\\
\vspace{0.45cm}
\includegraphics[width =0.45\textwidth]{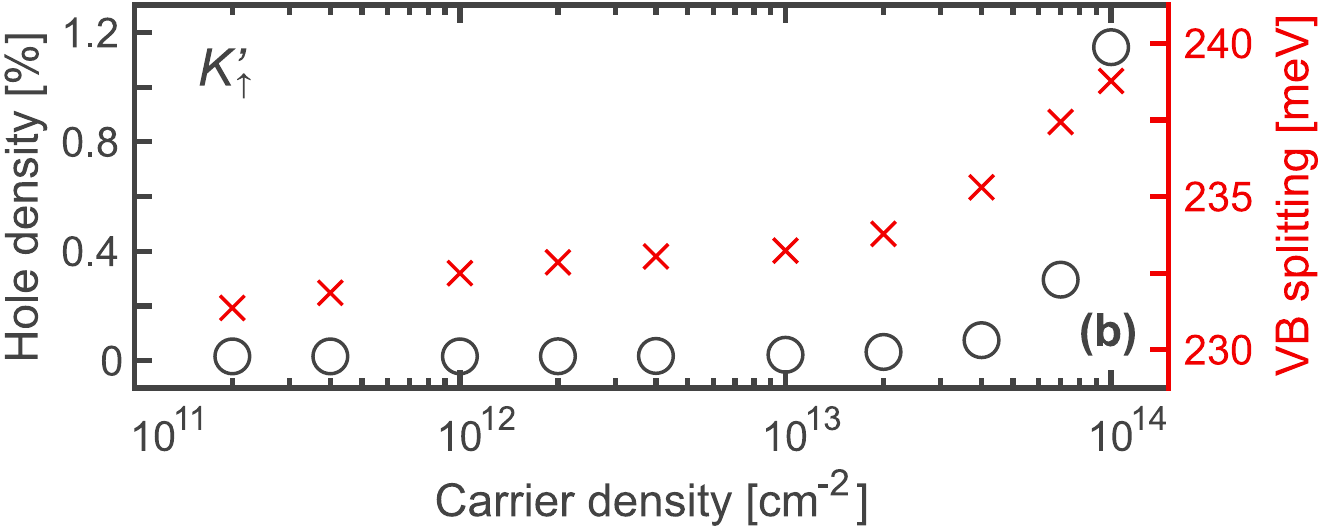}
\caption{\label{fig:7} (a) $\Sigma_{\uparrow}$ (top) and $\Lambda_{\uparrow}$ (bottom) side-valley electron densities (left axis, black circles) and $\Sigma_{\uparrow}$ (top) and $\Lambda_{\uparrow}$ (bottom) conduction-band offsets (right axis, red crosses) in dependence of the total amount of excited charge carriers. (b) Similarly, charge-carrier dependence of the $K'_{\uparrow}$ hole density (left axis, black circles) and $K_{\uparrow}/K'_{\uparrow}$ valence-band splitting (right axis, red crosses). The respective spin-up electron (hole) densities are stated as fraction of the total spin-up electron (hole) density. }
\end{figure}

The electron drain to the side valleys critically depends on the offset between the side valleys and the conduction band minima at the $K/K'$ points, which in turn are modified by the carrier occupations in each valley. In Fig.\,\ref{fig:7}\,(a), we depict the $\Sigma_{\uparrow}$ (top) and $\Lambda_{\uparrow}$ (bottom) side-valley electron densities (black circles) as well as the corresponding conduction-band offsets (red crosses) in dependence of the total density of excited charge carriers. Increasing the total carrier density from $2\times 10^{11}\,{\rm cm}^{-2}$ to $10^{14}\,{\rm cm}^{-2}$, the $\Sigma_{\uparrow}$ ($\Lambda_{\uparrow}$) conduction-band offset increases by $7.0$\,meV ($6.6$\,meV), while the fraction of $\Sigma_{\uparrow}$ ($\Lambda_{\uparrow}$) electrons increases from about $3.1$\,\% ($5.8$\,\%) in the low-density regime to about $8.8$\,\% ($15.1$\,\%) in the regime investigated by the pump-probe simulations. The larger electron drain towards the $\Lambda_\uparrow$ valleys arises from the smaller side-valley offset, that increases from $82$\,meV (DFT value) to $89.7$\,meV for the highest investigated carrier density. For the $\Sigma_\uparrow$ valleys, the band offset increases from $97$\,meV (DFT value) to $105.2$\,meV.

In addition to the electron drain, inter-valley scattering leads to a drift of the hole density between the $K$ and $K'$ valleys. In our calculations, the influence of this is taken into account in a similar manner as the side-valley drain, i.e. by considering an equilibrium situation with a common chemical potential for the holes in different valleys. In Fig.\,\ref{fig:7}\,(b), we plot the charge-carrier dependence of the fractional $K'_{\uparrow}$ hole density (black circles) and the corresponding $K_{\uparrow}/K'_{\uparrow}$ valence-band splitting (red crosses). Because of the large valence-band splitting  in the low-density limit, almost all the holes from the $K'_{\uparrow}$ valley have drifted to the energetically favorable $K_{\uparrow}$ valley after equilibration. Only for carrier densities as high as $10^{14}\,{\rm cm}^{-2}$, the $K'_{\uparrow}$ valley is occupied with at least $1$\,\% of the spin-up holes. Thus, the increasing valence-band splitting with increasing carrier density is exclusively introduced by the dominant occupation of holes in the $K_{\uparrow}$ valley. For the shown carrier densities, the valence-band splitting increases by $7.4$\,meV from $231.4$\,meV to $238.8$\,meV.

\begin{figure}[t!]
\includegraphics[width =0.45\textwidth]{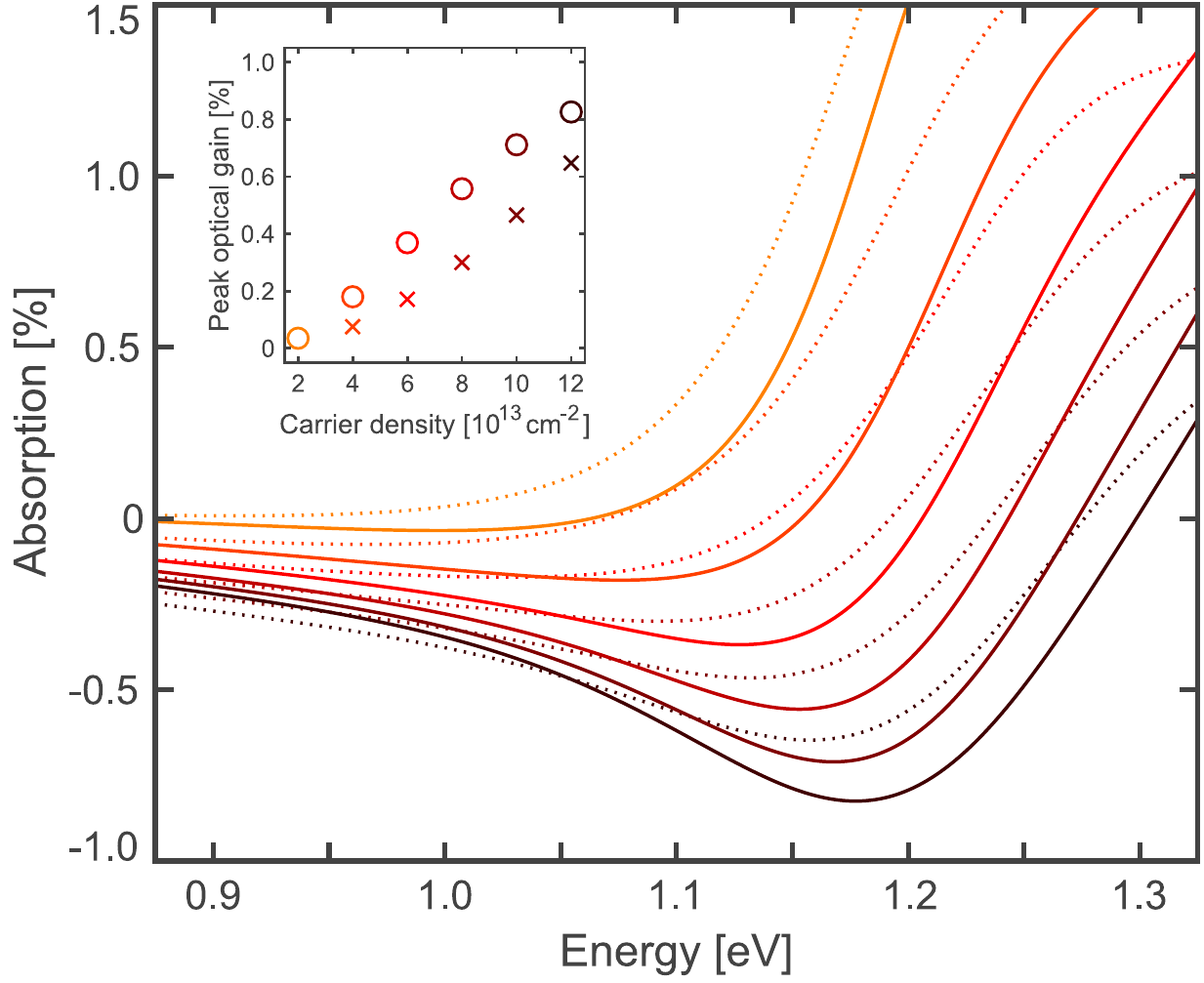}
\caption{\label{fig:8} Optical absorption spectra before and after equilibration of the carriers between the different valleys. The solid lines show the optical spectra after equilibration, the dotted lines before equilibration, where all carriers are located in the $K$ and $K'$ valleys. The inset shows the peak optical gain for the corresponding charge-carrier densities after (before) equilibration in circles (crosses).}
\end{figure}

Finally, we present the resulting optical absorption spectra after equilibration within the entire Brillioun zone for elevated charge-carrier densities in Fig.\,\ref{fig:8}. The carrier densities increase from orange ($2.0\times 10^{13}\,\mathrm{cm}^{-2}$) to black ($1.2\times 10^{14}\,\mathrm{cm}^{-2}$). The solid (dotted) lines show the optical spectra after (before) equilibration. The inset depicts the peak optical gain for the corresponding charge-carrier densities. Prior to equilibration, all carriers are assumed to be located in the $K$ and $K'$ valleys. As pointed out before, equilibration not only leads to electron drain to the side valleys, but also to hole drift between the $K$ and $K'$ valleys. Both processes have counteracting effects on the optical spectra. While the electron drain to the side-valleys leads to a loss of optically recombining electrons, the hole drift between the $K$ and $K'$ valleys increases the amount of optically recombining holes. Because of the large $K_{\uparrow}/K'_{\uparrow}$ ($K'_{\downarrow}/K_{\downarrow}$) valence-band splitting about $99$\,\% of the holes contribute to A-band population inversion after equilibration, compared to the nearly $50$\,\% before equilibration. This overcompensates the effect of electron drain -- between $10.8$\,\% and $26.9$\,\% of the electrons for the stated carrier densities -- to the side valleys. Consequently, enhanced peak optical gain is observed in equilibrium. In particular, for carrier densities of $0.8/1.0/1.2\times 10^{14}\,\mathrm{cm}^{-2}$ peak optical gain increases by $86/53/28$\,\% due to equilibration. For a carrier density as high as $1.2\times 10^{14}\,\mathrm{cm}^{-2}$, we observe peak optical gain occurring slightly below the low-density A-exciton resonance with a magnitude of $0.83\,\%$ of the incoming light. Normalized to the layer thickness of $D=6.99$\,\AA{} (see Table\,\ref{tab:DFTparameters}), this corresponds to a peak gain of about $10^5\,{\mathrm{cm}}^{-1}$, which should be compared to the gain maximum of $5000\,\mathrm{cm}^{-1}$ in typical III/V semiconductors under realistic excitation conditions\,\cite{nlcstr}.

\section{Discussion}

In summary, we investigated the carrier dynamics in ML \MoTe\ after excitation with a strong optical pump pulse slightly above the interacting $\romB$-band gap. Our investigations cover two distinct time regimes. In the excitation regime, i.e., during the optical pulse, generation of photo-induced charge carriers is accompanied by an almost instantaneous band-gap renormalization of about $410$\,meV in our case, that exceeds the exciton binding energy of the unexcited crystal. In the low-density regime, the band-gap renormalization is almost exactly canceled by the weakening of the excitonic binding such that the exciton resonance displays a negligible spectral shift. The initial fast carrier relaxation is followed by a much slower thermalization of the hot carriers. Due to efficient phonon coupling, the thermalization occurs within a few picoseconds, whereas it is typically of the order of tens of picoseconds in conventional III-V quantum well systems. For the chosen pump-pulse intensity, thermalization finally leads to population inversion. Here, we observe the transition from plasma absorption to broadband optical gain. The maximum of the A-band optical gain occurs slightly below the low-density A-exciton resonance and its magnitude approaches $0.5$\,\% of the incoming light. 

On the longer time scale, equilibration of the excited carriers among different valleys is expected with a simultaneous electron drain from the $K/K'$ to the side valleys at the $\Sigma/\Lambda$ points of the Brillioun zone. Although the electron drain leads to an efficiency drop of several percent in the gain regime, there is no evidence for an excitation-dependent roll-over from a direct to an indirect band gap, as has been predicted theoretically for similar material systems\,\cite{erben2018,lohof2019}. Instead, we find that the drop due to the electron drain is overcompensated by a hole drift between the $K$ and $K'$ valleys, leading to a net increase of the optical gain up to several $10$\,\% .

Our results are in general agreement with experimentally observed excitation-induced band-gap shrinkage {of similar TMDC systems\,\cite{chernikov2015,ulstrup2016}} and confirms the possibility of ultrafast band-gap modulation by the injection of carriers. Furthermore, our results identify conditions for achieving plasma gain {in ML \MoTe{}} and the short relaxation times enable high repetition rates.

\section{Acknowledgements}
The Marburg work of this project was funded by the DFG via the Collaborative Research Center SFB 1083. The Arizona work was supported by the Air Force Office of Scientific Research under award number FA9550-17-1-0246. We thank T.\ Sohier for  correspondence concerning the 2D Fr{\"o}hlich interaction in TMDC MLs.


%

\end{document}